\documentclass[apj]{emulateapj}
\newcommand\submitms{n}		
\if\submitms y
\else
\usepackage{apjfonts}
\fi

\usepackage{ifthen}
\usepackage{natbib}

\shorttitle{THE 1998 NOVEMBER 14 SATURN OCCULTATION.\ I.}
\shortauthors{HARRINGTON \& FRENCH}

\newcommand\degree{\degr}
\newcommand\degrees\degree
\newcommand\decdegree{\fdg}
\newcounter{fignum}

\DeclareSymbolFont{UPM}{U}{eur}{m}{n}
\DeclareMathSymbol{\umu}{0}{UPM}{"16}
\let\oldumu=\umu
\renewcommand\umu{\ifmmode\oldumu\else\math{\oldumu}\fi}
\newcommand\micro{\umu}
\renewcommand\micro{\math{\mu}}
\renewcommand\micron{\micro m}
\newcommand\microns \micron

\let\oldsim=\sim
\renewcommand\sim{\ifmmode\oldsim\else\math{\oldsim}\fi}
\let\oldpm=\pm
\renewcommand\pm{\ifmmode\oldpm\else\math{\oldpm}\fi}
\newcommand\by{\ifmmode\times\else\math{\times}\fi}
\newcommand\ttt[1]{10\sp{#1}}
\newcommand\tttt[1]{\by\ttt{#1}}
\newcommand\tablebox[1]{\begin{tabular}[t]{@{}l@{}}#1\end{tabular}}
\newbox{\wdbox}
\renewcommand\c{\setbox\wdbox=\hbox{,}\hspace{\wd\wdbox}}
\renewcommand\i{\setbox\wdbox=\hbox{i}\hspace{\wd\wdbox}}

\newcommand\herenote[1]{{\bfseries #1}\typeout{======================> note on page \arabic{page} <====================}}

\newcount\timect
\newcount\hourct
\newcount\minct
\newcommand\now{\timect=\time \divide\timect by 60
         \hourct=\timect \multiply\hourct by 60
         \minct=\time \advance\minct by -\hourct
         \number\timect:\ifnum \minct < 10 0\fi\number\minct}

\catcode`@=11

\if\submitms y
\renewcommand\comment[1]{}
\else
\newcommand\comment[1]{}
\fi
\newcommand\commenton{\catcode`\%=14}
\newcommand\commentoff{\catcode`\%=12}

\renewcommand\math[1]{$#1$}
\newcommand\mathshifton{\catcode`\$=3}
\newcommand\mathshiftoff{\catcode`\$=12}

\comment{the backslash is necessary}

\comment{alignment tab}

\let\atab=&
\newcommand\atabon{\catcode`\&=4}
\newcommand\ataboff{\catcode`\&=12}

\let\oldmsp=\sp
\let\oldmsb=\sb
\renewcommand\sp[1]{\ifmmode
	   \oldmsp{#1}%
	 \else\strut\raise.85ex\hbox{\scriptsize #1}\fi}
\renewcommand\sb[1]{\ifmmode
	   \oldmsb{#1}%
	 \else\strut\raise-.54ex\hbox{\scriptsize #1}\fi}
\newcommand\msp[1]{\ifmmode
	   \oldmsp{#1}
	 \else \math{\oldmsp{#1}}\fi}
\newcommand\msb[1]{\ifmmode
	   \oldmsb{#1}
	 \else \math{\oldmsb{#1}}\fi}
\newcommand\supon{\catcode`\^=7}
\newcommand\supoff{\catcode`\^=12}
\newcommand\subon{\catcode`\_=8}
\newcommand\suboff{\catcode`\_=12}
\newcommand\supsubon{\supon \subon}
\newcommand\supsuboff{\supoff \suboff}

\newcommand\actcharon{\catcode`\~=13}
\newcommand\actcharoff{\catcode`\~=12}

\newcommand\paramon{\catcode`\#=6}
\newcommand\paramoff{\catcode`\#=12}

\comment{And now to turn us totally on and off...}

\newcommand\reservedcharson{\commenton \mathshifton \atabon \supsubon \actcharon
	\paramon}

\newcommand\reservedcharsoff{\commentoff \mathshiftoff \ataboff
	\supsuboff \actcharoff \paramoff}

\newcommand\nojoe[1]{\reservedcharson#1\reservedcharsoff}

\catcode`@=12
\reservedcharsoff

\reservedcharson

\comment{ Must have ONLY ONE of these... trust these macros, they work

}
\reservedcharsoff

\actcharon
\if\submitms y
\newcommand\widedeltab{deluxetable}
\else
\newcommand\widedeltab{deluxetable*}
\received{2009 July 23}
\comment{\revised{2009 December 4}}
\accepted{2009 December 6}
\published{2010 May 19}
\fi

\if\submitms y
\slugcomment{\tablebox{
\comment{Submitted to {\em ApJ} Mon Oct  4 21:52:00 EDT 2004.}
Submitted to {\em ApJ} 2009 July 23.\\
\comment{Revised 2009 December 4.\\}
Accepted 2009 December 7.}}
\else
\slugcomment{Version of 2010 May 19 for arXiv.org.}
\journalinfo{{\em To appear in} The Astrophysical Journal, 716:398--403, 2010 June 10
             \hfill {\em as} {\rm doi}:10.1088/0004-637X/716/1/398}
\fi

\begin{document}

\title{The 1998 November 14 Occultation of GSC 0622-00345 by Saturn.\\
  I. Techniques for Ground-Based Stellar Occultations}
\author{Joseph Harrington\altaffilmark{1,3}}
\author{Richard G.\ French\altaffilmark{2,3}}
\affil{\sp{1}Planetary Sciences Group, Department of Physics,
  University of Central Florida, Orlando, FL 32816-2385, USA;
  jh@physics.ucf.edu}
\affil{\sp{2}Astronomy Department, Wellesley College, Wellesley, MA
  02481, USA; rfrench@wellesley.edu}
\altaffiltext{3}{Visiting Astronomer at the Infrared Telescope
  Facility, which is operated by the University of Hawaii under
  Cooperative Agreement no.\ NCC 5-538 with the National Aeronautics
  and Space Administration, Science Mission Directorate, Planetary
  Astronomy Program.}

\begin{abstract}

\noindent On 1998 November 14, Saturn and its rings occulted the star GSC
0622-00345.  We observed atmospheric immersion with NSFCAM at the
National Aeronautics and Space Administration's Infrared Telescope
Facility on Mauna Kea, Hawaii.  Immersion occurred at 55{\decdegree}5 S
planetocentric latitude.  A 2.3 {\micron}, CH\sb{4}-band filter
suppressed reflected sunlight.  Atmospheric emersion and ring data
were not successfully obtained.  We describe our observation,
light-curve production, and timing techniques, including improvements
in aperture positioning, removal of telluric scintillation effects,
and timing.  Many of these techniques are known within the occultation
community, but have not been described in the reviewed literature.  We
present a light curve whose signal-to-noise ratio per scale height is
267, among the best ground-based signals yet achieved, despite a
disadvantage of up to 8 mag in the stellar flux compared to prior
work.

\if\submitms y
\else
\comment{\hfill\herenote{DRAFT of {\today} \now}.}
\fi
\end{abstract}

\section{INTRODUCTION}
\label{intro}

Stellar occultations by planets offer among the highest spatial
resolution of any astronomical observing technique, competing even
with atmospheric descent probes for certain measurements.  From Earth,
spatial resolution is limited by the projected stellar beam size and
Fresnel diffraction, the latter giving \sim1 km resolution for solar
system giant planets for visible and near-IR wavelengths.  This is
sufficient to reveal fine structure in a planetary ring or propagating
gravity (buoyancy) waves in an atmosphere.  However, the occultation
technique is demanding, and failing to optimize even a single
parameter of an observation can significantly reduce the quality of
the derived light curve.  Occultation observers have thus developed
numerous specialized techniques to improve their data.  Many of these
have never been described in the literature.

On 1998 November 14, Saturn and its rings occulted GSC 0622-00345, as
predicted by \citet{BoshMcDonald1992ajsatocs}.  We obtained a
light curve for atmospheric immersion, based on infrared imaging
observations at the NASA Infrared Telescope Facility (IRTF) on Mauna
Kea, HI.  Figure \ref{geometry} shows our viewing circumstances and
Table \ref{occtab} gives parameters of the event.

Occultation cadences are typically from a few to 10 Hz, and in the case of
lunar occultations they can be much higher.  These rates are
sufficient for the temporal resolution of flux variations caused by
scintillation (variable refractive focusing and defocusing of the
stellar beam) in Earth's atmosphere.  This is typically the
limiting noise source for occultations by bright stars.  During our
observations, we monitored Rhea, which experienced approximately the
same scintillation effects as the star, just arcseconds away.  We were
thus able to use measurements of Rhea's (assumed constant) flux to
compensate for scintillation.  This, in combination with other
techniques, produced a light curve of very high quality.

Here we present a full accounting of the methods we employed to
observe the event and derive the light curve, including those
previously undocumented.  Section \ref{obslcsec} describes observing
techniques, data reduction methods, and the light curve.  It also
discusses an attempt to apply optimal photometry to light curve
production and thoughts on improving instrument interfaces to avoid
failures such as our lost egress observation.  Section \ref{timingsec}
presents a method for independently verifying the system timing
solution.  Section \ref{conclsec} offers brief conclusions.  We
include the light curve and software for analyzing occultation timing
data as electronic supplements.  Paper II
\citep{HarringtonEtal2010apjSatoc98II} uses the derived light curve to
investigate the atmosphere of Saturn.

\section{OBSERVATIONS AND LIGHT CURVE}
\label{obslcsec}

\begin{figure}[th]
\if\submitms y
  \setcounter{fignum}{\value{figure}}
  \addtocounter{fignum}{1}
  \newcommand\fignam{f\arabic{fignum}.eps}
\else
  \newcommand\fignam{JH_occgeom_rev4.ps}
\fi
\plotone{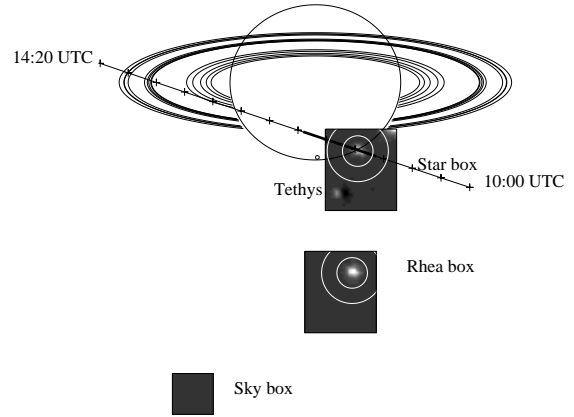}
\figcaption{\label{geometry}
Occultation geometry, subarray placement, and photometry apertures at
the half-light time.  The light line marks the path of GSC 0622-00345
relative to Saturn, with crosses every 20 minutes.  The heavier
portion gives the span of the data set.  The relative positions of the
star, Rhea, and sky subarrays were constant.  Telescope tracking kept
the star in roughly the same pixel while Saturn and its moons drifted
through.  Tethys's motion relative to Saturn is evident by the
position mismatch between the half-light data and the scattered-light
template.  We computed the template from post-occultation frames and
subtracted it from the data to produce the star frame shown here.
}
\end{figure}

The NSFCAM InSb array camera \citep{ShureEtal1994exansfcam} recorded
the event in ``movieburst'' mode.  This mode writes a time sequence of
images to computer memory and then saves the set to disk after the
observations finish.  We recorded three subarrays derived from the
full 256 {\by} 256 pixel array.  These boxes contained the star
(28 {\by} 32 pixel), Rhea (28 {\by} 32 pixel), and a piece of blank sky
(16 {\by} 16 pixel), placed as shown in Figure \ref{geometry}.  The
larger boxes were originally 32 {\by} 32 pixels, but a camera software
error overwrote four edge columns.  No critical data were lost and the
error is now fixed.  Table \ref{lctab} gives additional observation
and light-curve parameters.

The limit to spatial resolution in a stellar occultation is generally
the larger of the Fresnel scale and the projected size of the occulted
star.  For typical in situ spacecraft occultations, both are
very small compared to the sky-plane distance between samples, but for
Earth-based work either can dominate.  In the case of 28 Sgr, the
projected stellar diameter at the distance of Saturn was \sim20 km,
much larger than the \sim1-km Fresnel diffraction scale, resulting in
significant smoothing of the ring and atmospheric occultation profiles
\citep[e.g.,][Figure 4]{HarringtonEtal1993ic28sgr}.  To estimate the
angular diameter of GSC 0622-00345, we first used the color index
\math{V - K = 2.607} (Table \ref{occtab}) to infer a dwarf spectral
type of K5 \citep{Glass1999bookHIA}.  The stellar diameter for this
spectral class \citep{Lang1992bookADPS} yields an estimated angular
diameter of 0.12 mas, or 0.7 {\pm} 0.2 km at Saturn's distance, where
the error bar takes into account the possible effects of modest
interstellar reddening.  Even with a large uncertainty, the projected
size of the occulted star is less than the Fresnel scale, which thus
determines the intrinsic spatial resolution of the occultation.

The 2.3-{\micron} filter is centered on a methane band, where Saturn's
atmosphere strongly absorbs sunlight, so contrast between the star and
Saturn is very high.  The rings are bright at this wavelength, but
their spatial separation from the immersion latitude made it practical
to subtract their scattered light (see below).

\atabon\begin{deluxetable}{l@{\extracolsep{0.25em}}r@{\extracolsep{0.25em}}l@{\extracolsep{0.25em}}l}
\tablecaption{\label{occtab} Occultation Parameters}
\tablewidth{0pt}
\tablehead{
\colhead{Description} &
\multicolumn{2}{c}{Value} &
\colhead{Comment}}
\startdata
Star				& \multicolumn{3}{l}{GSC 0622-00345} \\
\hspace{2 em}R.A.		& 1h49m54{\farcs}358 & & J2000\tablenotemark{a} \\
\hspace{2 em}Decl. 		& 8{\degrees}23{\arcmin}12{\farcs}56 && J2000\tablenotemark{a} \\
\hspace{2 em}\math{V} magnitude	& 11.002 & & \tablenotemark{a} \\
\hspace{2 em}\math{K} magnitude	& 8.395 & & \tablenotemark{b} \\
\hspace{2 em}Spectral Type	& K5 & & \\
\hspace{2 em}Proj.\ diam.\ at Saturn & \sim0.7 & km \\
Half-light latitude		& 55{\decdegree}5 & S & Planetocentric \\
Fresnel scale			& 1.2 & km & \math{\sqrt{\lambda d / 2}} \\
Sky-plane velocity		& 17.940 & km s\sp{-1} \\
Perpendicular velocity, \math{v\sb{\perp}} & 12.802 & km s\sp{-1} & vs.\ oblate limb \\
Ring opening angle		& -14{\decdegree}6772 \\
Pole position angle		& 1{\decdegree}5134 &  & J2000 \\
Saturn's geocent.\ dist., \math{d} & 1.25168\tttt{9} & km
\enddata
\tablenotetext{a}{Tycho-2 record 131066, \citet{HogEtal2000aaTycho2},
  \math{\sigma\sb{\rm pos} \sim 65} mas, \math{V = VT - 0.090 (BT -
    VT)}, using catalog \math{VT} and \math{BT} magnitudes}\newline
\tablenotetext{b}{2MASS 01495435+0823123,
  \citet{SkrutskieEtal2006AJ2MASS}}
\end{deluxetable}\ataboff
\placetable{occtab}

At the half-light time (see below), seeing was 0{\farcs}9 FWHM and the
star was at 1.3 air masses and setting.  The night was clear, but
Rhea's flux varied by up to 8
scintillation.  Rhea was nearly as bright as the unocculted star (see
Table \ref{lctab}) and had a point-spread function (PSF) that very
closely matched that of the star in each frame.  We thus used Rhea as
a standard for flux and position, and attempted to use it as a PSF
standard.  In contrast to the scintillation, sky emission was very
steady and low, based on the separately recorded sky box.

To obtain sub-pixel accuracy in image shifts, photometry aperture
placement, and scattered-light subtraction, we subsampled by a factor
of 10 using nearest-neighbor sampling (a factor of 5 gave essentially
identical results).  We performed all operations on the expanded
images, except as noted below.  These subpixels allowed a
reasonable approximation to a circular photometry aperture that
included partial pixels.

\if\submitms y
\clearpage
\fi
\if\submitms y
\clearpage
\fi

After debiasing, flat-fielding, and interpolating three hot pixels
that were not near the star or Rhea, we centroided Rhea in all images.
The relative position of the star in two well-separated sets of 100
pre-occultation frames determined the Rhea-star offset and the
offset's drift with time.  This allowed accurate placement of the
Saturn and ring background template (described below) and photometry
apertures, even when the star was fully occulted.  A third set of
stellar centroids confirmed that the predicted stellar positions were
accurate to within 0{\farcs}045 (rms), or 0.15 unscaled pixels.  To
estimate Saturn's position (for template placement, see below), we added
the Rhea-Saturn relative sky-plane offset to Rhea's measured position
in each frame.  The Jet Propulsion Laboratory's Horizons On-Line
Ephemeris System
Web site\footnote{http://ssd.jpl.nasa.gov/horizons.html} generated the
relevant positions on 2004 January 9 based on the DE-0406, LE-0406,
and SAT136 ephemerides.  Our derived Rhea-star drift rate closely
matched the ephemeris prediction.

\if\submitms y
\clearpage
\fi
\atabon\begin{\widedeltab}{lr@{\extracolsep{0.25em}}ll}
\if\submitms y
\tabletypesize{\scriptsize}
\fi
\tablecaption{\label{lctab} Observation and Light-curve Parameters}
\tablewidth{0pt}
\tablehead{
\colhead{Description} &
\multicolumn{2}{c}{Value} &
\colhead{Comment}}
\startdata
Telescope			& IRTF \\
Instrument			& NSFCAM \\
Image scale			& 0{\farcs}30 & pixel\sp{-1} \\
Wavelength, \math{\lambda}	& 2.28 & \microns & ``Spencer 2.3'' filter \\
Bandpass, FWHM			& 0.17 & \microns \\
Time between frame starts	& 0.2128 & s \\
Exposure time			& 0.1276 & s & Dead time of 0.0852 s, see the text \\
Readout time                    & 0.009469 & s \\
Non-destructive reads (NDR)     & 8 \\
Gain                            & 10 & e\sp{-} DN\sp{-1} \\
Read noise                      & 19 & e\sp{-} & 55 e\sp{-} read\sp{-1} / \math{\sqrt{{\rm NDR}}}\\
\multicolumn{4}{l}{Frame mid-times:} \\
\hspace{2 em} First frame	& 11:05:02.622 & UTC \\
\hspace{2 em} First frame in upper baseline average	& 11:16:44.939 & UTC \\
\hspace{2 em} First frame in lower baseline average	& 11:26:19.562 & UTC & No evidence of star in frames \\
\hspace{2 em} Last frame	& 11:56:07.065 & UTC \\
Number of frames		& 14,400 \\
Number of frames per baseline average & 300 \\
\multicolumn{4}{l}{Photometry (upper baseline interval):} \\
\hspace{2 em} Star full flux	& 5439 & DN & Above lower baseline \\
\hspace{2 em} Rhea full flux	& 5216 & DN & Above lower baseline \\
\hspace{2 em} Sky box		& 12.55 & DN pixel\sp{-1} \\
\multicolumn{3}{l}{Variability (upper baseline interval):} 
                                          & Standard deviation \\
\hspace{2 em} Star		& 2.2 & 
\hspace{2 em} Star		& 1.5 & 
\hspace{2 em} Sky box		& 3.8 & 
S/N per frame			& 66 \\
S/N per scale height		& 267
\enddata
\end{\widedeltab}\ataboff
\if\submitms y
\clearpage
\fi
\placetable{lctab}

The small amount of scattered light from Saturn and the rings
complicated stellar photometry by requiring the subtraction of a
scattered-light template.  We made a template by calculating the
shifted, masked average (see below) of a set of post-immersion images,
each divided by Rhea's flux in that frame and shifted to align the
predicted position of Saturn in all frames.  All shifts both in
creation and application of the template were in units of integer
subpixels, so no interpolation occurred.

A shifted, masked average starts with two similar 3D ``stacks,''
respectively containing the images and their bad pixel masks (in our
case, all the masks are the same).  In a mask, pixel values are one
for good pixels and zero for bad pixels.  The technique proceeds as
follows: pad the stacks with zeros to prevent good data from shifting
outside the boundaries; multiply the image stack by the mask stack to
set bad pixels to zero, shift the images to align them, and apply the
same shifts to the masks; collapse each stack into a 2D image by
summing along the image index dimension; at each pixel location,
divide the image sum by the mask sum to calculate the shifted mask
average.  Each stacked image shifted differently, so the bad pixels
shifted out of alignment with one another.  The result has no bad
pixels and was calculated without interpolation.  The mask sum gives
the number of good pixels contributing to the average at each pixel.

To apply the template, we shifted it according to Saturn's predicted
position (see above) in each occultation frame, multiplied it by
Rhea's flux in that frame, down-sampled to the original spatial
resolution, subtracted from the calibrated stellar frame, up-sampled
the result to the scaled resolution using nearest-neighbor sampling,
and performed aperture photometry using the predicted stellar
position.

The 0.15-pixel centering uncertainty would lead to slight
misalignments of the template.  In the region of the limb, the
template's gradient is typically 0.0018 DN per subpixel (the peak is
0.0054 DN per subpixel).  Both these values are found after multiplying by
Rhea's mean flux from Table \ref{lctab} and dividing by the square of
the scaling (100, so that photometry yields an unscaled DN value like
those in Table \ref{lctab}).  The aperture boundary touches the limb
gradient in only two places, however; so template shift errors
contribute minimal noise.  Shifting the template by one or two
subpixels horizontally or vertically, subtracting it from the
unshifted template, and performing aperture photometry yields -1.26
and -2.10 DN for the horizontal shifts and 1.21 and 2.08 DN for the
vertical shifts, respectively.  The photon noise from the unocculted
star is 23 DN, from the sky in the aperture is 12 DN, and from other
sources is 15 DN based on aperture photometry of the template.
Template misalignment is thus a minor noise source even compared to
photon noise.

The template-subtracted stellar frames were quite flat except in the
corner that included some ring light (see Figure \ref{geometry}).
Pixels in this region were in the sky annulus but not in the
photometry aperture, and were efficiently rejected by taking the
median of all good pixels in the sky annulus.  Sky photometry for Rhea
also used the median, for consistency, as did that in the preceding
paragraph.

The stellar and Rhea photometry apertures both had diameters of
3{\farcs}6, four times the seeing FWHM.  The abutting sky annuli had
outer diameters of 7{\farcs}2 (see Figure \ref{geometry}).  The
photometry for the unocculted star and for Rhea closely followed each
other, indicating that the main noise contributor was scintillation.
We thus divided the stellar flux by Rhea's flux in each frame (just as
we adjusted the template's flux).

\if\submitms y
\clearpage
\fi
\begin{figure*}[th]
\if\submitms y
  \setcounter{fignum}{\value{figure}}
  \addtocounter{fignum}{1}
  \newcommand\fignam{f\arabic{fignum}.eps}
\else
  \newcommand\fignam{plot_isofit_2005_04_15a.ps}
\fi
\centerline{
\includegraphics[angle=90, width=0.8\textwidth, clip]{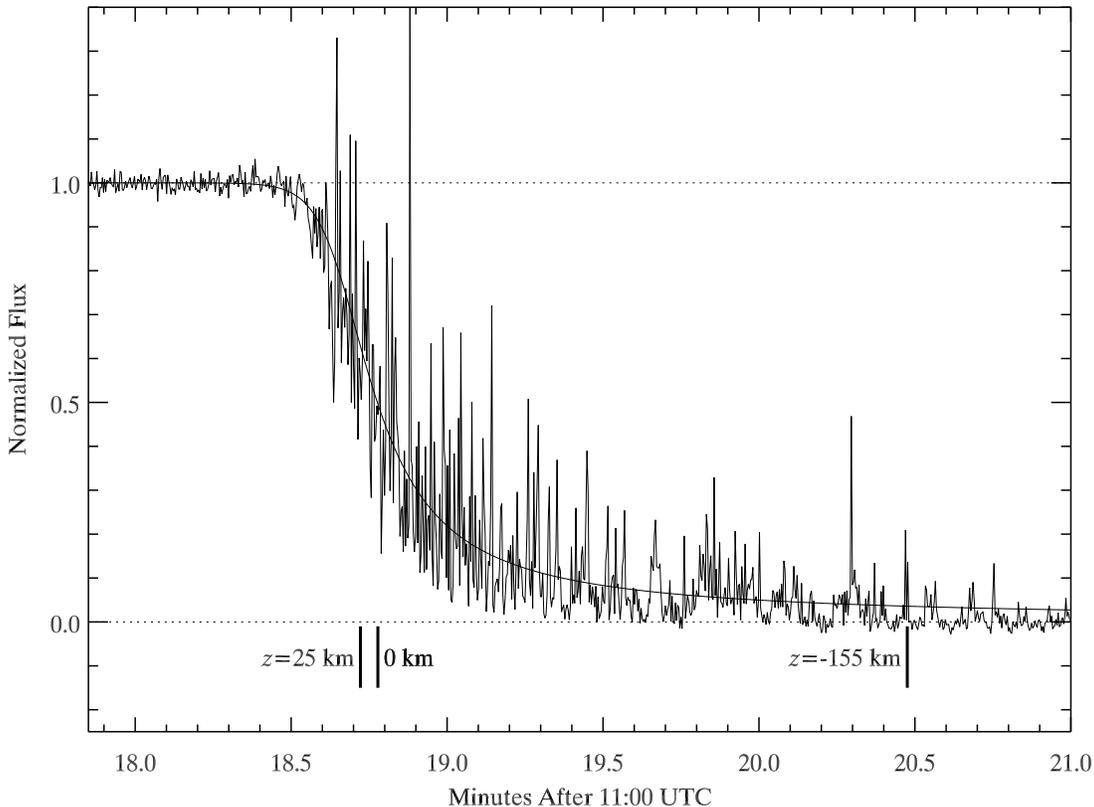}
}
\figcaption{\label{lightcurve}
Light Curve for occultation of GSC 0622-00345 by Saturn, observed at
the NASA IRTF on 1998 November 14 UTC.  The light curve has stable
upper and lower baselines (dashed lines) and quite low photometric and
terrestrial atmospheric scintillation noise.  This makes it possible
to determine the thermal profile of the atmosphere by numerical
inversion.  The light curve is characterized by many sharp, narrow
spikes, due to small-scale refractive focusing by density fluctuations
in Saturn's atmosphere.  In some cases, these spikes have an amplitude
of several times the ambient intensity in the light curve.  The smooth,
solid line is an isothermal model fit with baselines fixed (see Table
1 of Paper II for parameters), which we note in the text does not fit
well near the lower baseline.  Below the light curve, we mark the
location of the half-light point (\math{z = 0} km) and the extrema of
the reliable region of the inversion, using altitudes determined by
the inversion.
}
\end{figure*}
\if\submitms y
\clearpage
\fi

The light curve's upper and lower baseline levels are the medians of
sections of the data before and after immersion (see Table
\ref{lctab}).  We used the median to reduce the effect of non-Gaussian
residual scintillation.  The lower baseline estimate and the template
come from the same frames.  Verifying that there is no star in the
lower-baseline frames is crucial to the fits and inversions that
follow.  The lower baseline frames were aligned at the nominal Saturn
position and added to make the template.  Due to increasing ray
bending, the nominal stellar position is a relatively stationary point
on the limb at this time in the occultation.  The template shows no
evidence of the star and is smooth (see above).  After template
subtraction and scintillation correction, the median integrated value
for the star aperture in the lower-baseline frames was (1.5 {\pm}
6.0)\tttt{-4} units of the full stellar flux over the level in the sky
annulus.  The mean was less than twice this value, and the uncertainty
is the standard deviation of the mean.  The star is effectively gone.

Figure \ref{lightcurve} presents the light curve.  Its per-frame
signal-to-noise ratio (S/N)
is 66 for the unocculted star.  The
\citet{FrenchEtal1978icocinv} atmospheric occultation noise parameter
\math{\epsilon\sb{\phi} = \epsilon\sqrt{v\sb{\perp}/H}} = 0.00375,
where \math{\epsilon} is the standard deviation of the unocculted
stellar flux in one-second bins divided by its mean,
\math{{v\sb{\perp}}} is given in Table \ref{occtab}, and \math{H}, the
scale height, is given in Table 1 of Paper II (baselines fixed case).
The reciprocal of \math{\epsilon\sb{\phi}} is the S/N per scale
height, and both are commonly used as figures of merit for occultation
light curves.  Our value of 267 exceeds even that of the extremely
bright 28 Sgr occultation.  For example, the IRTF 28 Sgr immersion S/N
per scale height was 192, using the fractional flux standard deviation
per frame of 1.7
\citet{HarringtonEtal1993ic28sgr} and \math{H/v\sb{\perp}} = 2.66
reported by \citet{HubbardEtal1997icsatmes}.  The raw S/N for these
occultations is limited by scintillation rather than by photon
statistics, eliminating 28 Sgr's 8 mag brightness advantage
over the event reported here.  Our slightly slower event, improvements
in infrared array technology, and the analysis methods described
herein account for the rest.  The electronic supplement to this
article includes the light curve data file.

A histogram of our upper baseline residuals closely follows a Gaussian
distribution to \math{\pm3\sigma}, but at \math{>}+3\math{\sigma} it
is slightly above the Gaussian, perhaps because scintillation produces
occasional, transient flashes.  The standard deviation in the
difference between adjacent-frame fluxes is \math{>}93
expected for uncorrelated errors, and reaches \sim99
of 1 s.  The residual correlation may come from the \math{<5} ms
interval between reads of the star and Rhea pixels.  The power
spectrum of the noise is thus still slightly stronger at lower
wavenumbers than at higher wavenumbers (see below).  In the
occultation itself, one would expect correlated structure on the scale
of the projected stellar diameter or the Fresnel diffraction scale
(see Table \ref{occtab}), but both of these are below the 2.7-km
sky-plane resolution of the light curve.

We computed the nominal occultation track from the ephemeris, and then
offset by an additional 500 km E and 500 km N, relative to Saturn's
center, so that the absolute radius of the half-light level matched
the oblate half-light surface determined by
\citet{HubbardEtal1997icsatmes} from 28 Sgr occultation observations.
This shift is within the astrometric accuracy of the star relative to
Saturn.

Since the PSFs of Rhea and the star were nearly the same, we attempted
to perform optimal stellar photometry, analogous to the optimal
spectral extraction of \citet{Horne1986paspoptextr} and others.  One
of us (J.H.) used this method for exoplanetary secondary eclipses from
the {\em Spitzer Space Telescope}\/ {\citep{DemingEtal2005natHD209458b}}, but
developed it initially for the present analysis.  We divided each
background-subtracted point-source image by a normalized PSF template.
Each pixel in the subarray thus became an estimate of the total flux
of the point source, with a variance that increased (rapidly) with
distance from the stellar centroid.  We then computed an
error-weighted (i.e., optimal) mean of all the good pixels in the
frame.

In principle, this process can reduce the noise by up to 70
optimal weighting discounts pixels far from the stellar centroid, and
thus reduces the effect of read noise and residual photon and
scintillation noise from scattered planetary light.  There is also no
photometry aperture size decision to make.  However, the method
requires an accurate PSF model and good centroids.  The highly varying
and non-Gaussian PSFs of these short-exposure images precluded any
smoothing or analytic PSF model fit, so the PSF templates were just
the background-subtracted, normalized images of Rhea, shifted to the
centroided position of the star in its images and down-sampled to the
original resolution.  For both Rhea and the star, we used the
background level from aperture photometry.

Regrettably, while this procedure yielded a lower baseline that was
three times less noisy than in the aperture-derived light curve, the
upper baseline was about four times noisier.  Close examination of the
PSFs showed that noise and the fractional-pixel centroiding errors
were sufficient to compromise this method.  We decided, for
the sake of consistency, not to merge the initial aperture photometry with
the final optimal photometry.  However, a much brighter PSF standard
might have substantially improved our photometry.  Obtaining such a
standard may be difficult in occultation experiments with bright
stars, but we note that large satellites could still play this role in
occultations of fainter stars, potentially increasing the number of
useful occultation events.  If optimal photometry were successful, the
disappearance of scintillation noise with atmospheric depth would
argue for the application of significance tests similar to the scaled
noise tests of Paper II, as opposed to the unscaled set.

We did not acquire useful egress data.  Atmospheric emersion occurred
in the rings (see Figure \ref{geometry}), so, following
\citet{HarringtonEtal1993ic28sgr}, we employed a 3.34 {\micron} filter
(wavelength measured warm) in which both planet and rings are dark due
to overlapping absorption bands of water ice and methane.  To increase
the frame rate, NSFCAM's movieburst mode cannot display data during
acquisition.  It was thus not possible to see until afterward that
the frames were empty.  Possible reasons include a rise in the
telluric water column (which would absorb all the light at this
wavelength), a filter wheel problem, or (unlikely) a pointing problem.
Two processes using shared memory on a dual-CPU camera computer could
implement a movieburst mode with simultaneous image display without
compromising the frame rate.

\section{TIMING AND ARRAY CLOCKING}
\label{timingsec}

Relating each flux measurement to a location in the atmosphere
requires precise knowledge of the time of each exposure.  At the start
of the first movieburst frame, NSFCAM's computer reads the time from a
Global Positioning System (GPS) receiver.  The observer must find the
delay between the GPS read and the actual first collection of photons,
the actual exposure time, and the time to read the array (which is
done many times per frame).  For this purpose, there is a flashing
light-emitting diode (LED) that is driven by the GPS and located in
the camera.  The LED turns on for 0.1 s at the start of each
Coordinated Universal Time (UTC)
second.  It is often lit for only part of an exposure, resulting in
count levels that are between the fully lit and unlit levels.  By
observing this LED with exactly the same settings as for the
occultation, one can model the count levels to determine the timing
parameters.

\if\submitms y
\clearpage
\fi
\begin{figure}[t]
\if\submitms y
  \setcounter{fignum}{\value{figure}}
  \addtocounter{fignum}{1}
  \newcommand\fignam{f\arabic{fignum}.eps}
\else
  \newcommand\fignam{satoc98-imtiming.ps}
\fi
\includegraphics[angle=90, width=\columnwidth, clip]{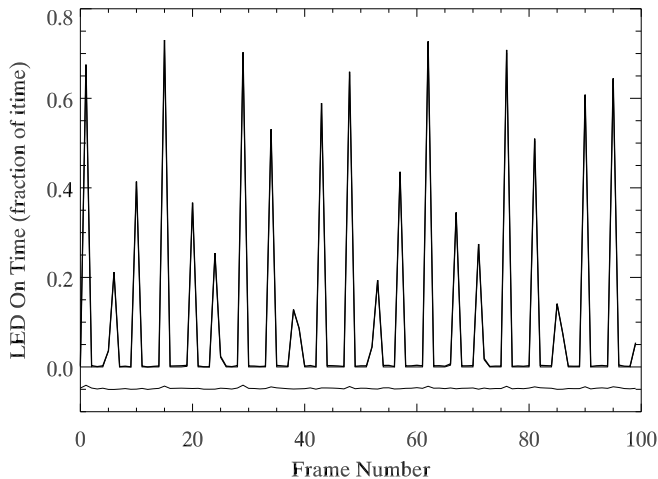}
\figcaption{\label{timing}
Portion of the LED timing light curve and the best-fit model.  Traces
for both the LED light curve and the model overplot each other at the
top of the plot.  The lowest trace presents the residuals, offset by
-0.05 for clarity.
}
\end{figure}
\if\submitms y
\clearpage
\fi

We made such measurements before and after the occultation.  For each
frame, we summed the flux in the sky subarray (which was lit by the
LED) after subtracting from each pixel the average per-pixel flux in
the star subarray (which was not).

Our model starts with five lists of event times: frame resets,
pedestal non-destructive reads (NDRs, see below), readout NDRs, LED
``on'' times, and LED ``off'' times.  It concatenates the lists, sorts
them in time, and steps through the result to calculate the flux per
frame.  It thus takes into account the changes in effective
sensitivity induced by each NDR.  The five free parameters are delay
after GPS read, integration time, readout time, and the two LED flux
levels.  We iteratively fit the model to the LED data by minimizing
the rms residuals.  The relevant results appear in Figure \ref{timing} and
Table \ref{lctab}.  The electronic supplement to this article includes
code that implements the model in the Interactive Data Language
(a product of ITT Visualization Systems, Boulder, CO, USA).

The per-image timing uncertainty (``jitter'') is difficult to
estimate, as fluctuations in the LED brightness, read noise, etc.\ all
masquerade as jitter.  This parameter is important in studies of
detailed light curve structure, such as the shape and location of
spikes, but this is not the focus of our present investigation.

Future instruments will likely read the GPS for each frame.  However,
an end-to-end test such as that provided by NSFCAM's LED builds
confidence that these increasingly-sophisticated instruments have
consistent timing.  It also quantifies the actual integration time and
the delay between the GPS read and the first detected photons.

We used NSFCAM's default ``array reset clocking, double sample'' mode.
This mode performs a series of NDRs both
before and after the exposure, and records the difference between the
two sets' averages.  The multiple NDRs suppress read noise but reduce
sensitivity for part of the exposure.  We used the NSFCAM default of
eight NDRs, but should have reduced dead time and increased read noise
by using fewer NDRs.  There were two effects of our long dead time.
First, the photon noise was higher than it could have been.  Since
scintillation still dominated the noise even after our correction, this is
regrettable but not serious.  Second, flux variations occurring during
either set of NDRs were only fractionally recorded, depending on when
they occurred in the set.  We may have missed some of the light in fast
spikes.  These correspond to the smallest-scale features in the
inversion presented in Paper II.  Since the small scales have the
smallest amplitudes (as expected, see Wavelet Analysis, Paper II), we
conclude that our error did not strongly affect the results.

\section{CONCLUSIONS}
\label{conclsec}

We have presented a light curve based on IRTF observations of the 1998
November 14 occultation of GSC 0622-00345 by Saturn.  The light curve
has a per-frame S/N of 66 and a per-scale-height S/N of 267, placing
it among the best ground-based atmospheric light curves.  We have
described the observational and analytic methods used to derive the
light curve, in some cases for the first time in the literature,
although these techniques are known among experienced occultation
observers.  Methods that improved the photometry include using Rhea as
a scintillation and pointing standard, spatially subsampling the data
by a factor of 10 for accurate aperture placement, and removing a
template of scattered light.  Separate, identical, measurements of a
flashing LED placed in the telescope beam provided an independent
measure of the frame rate, when fit by a detailed model of the chip's
sensitivity throughout the imaging cycle.  Optimal photometry shows
promise at reducing the photometric noise level as the star dims, but
requires a PSF standard that is substantially brighter than the
occultation star.  Occultation cameras with wider fields of view (but
still able to operate with accurate timing at high frame rates) would
facilitate finding bright PSF standards.

\acknowledgments

We thank P.\ Nicholson for his pre-occultation calculations and
J.\ L.\ Elliot, E.\ Lellouch, and L.\ Young for comments that
significantly improved the paper.  We thank the NASA Astrophysics Data
System, JPL Solar System Dynamics group, and the free and open-source
software communities for software and services.  This investigation
was supported by Wellesley College under NASA Contract 961169 and by
the NASA PGG program.

{\em Facilities:\/} IRTF

\nojoe{
\if\submitms y
  \newcommand\bblnam{ms}
\else
  \newcommand\bblnam{satoc98-1}
\fi
\bibliography{\bblnam}}

\begin{thebibliography}{12}
\expandafter\ifx\csname natexlab\endcsname\relax\def\natexlab#1{#1}\fi

\bibitem[{{Bosh} \& {McDonald}(1992)}]{BoshMcDonald1992ajsatocs}
{Bosh}, A.~S., \& {McDonald}, S.~W. 1992, \aj, 103, 983

\bibitem[{{Deming} {et~al.}(2005){Deming}, {Seager}, {Richardson}, \&
  {\jhauth{Harrington}}}]{DemingEtal2005natHD209458b}
{Deming}, D., {Seager}, S., {Richardson}, L.~J., \& {\jhauth{Harrington}},
  {\jhauth{J}}. 2005, \nat, 434, 740, astro-ph/0503554

\bibitem[{{French} {et~al.}(1978){French}, {Elliot}, \&
  {Gierasch}}]{FrenchEtal1978icocinv}
{French}, R.~G., {Elliot}, J.~L., \& {Gierasch}, P.~J. 1978, Icarus, 33, 186

\bibitem[{{Glass}(1999)}]{Glass1999bookHIA}
{Glass}, I.~S. 1999, {Handbook of Infrared Astronomy} (Cambridge: Cambridge
  Univ.~Press)

\bibitem[{{Harrington} {et~al.}(1993){Harrington}, {Cooke}, {Forrest},
  {Pipher}, {Dunham}, \& {Elliot}}]{HarringtonEtal1993ic28sgr}
{Harrington}, J., {Cooke}, M.~L., {Forrest}, W.~J., {Pipher}, J.~L., {Dunham},
  E.~W., \& {Elliot}, J.~L. 1993, Icarus, 103, 235

\bibitem[{{Harrington} {et~al.}(2010){Harrington}, {French}, \&
  {Matcheva}}]{HarringtonEtal2010apjSatoc98II}
{Harrington}, J., {French}, R.~G., \& {Matcheva}, K. 2010, \apj, 716, 404,
  (Paper II)

\bibitem[{{H{\o}g} {et~al.}(2000){H{\o}g}, {Fabricius}, {Makarov}, {Urban},
  {Corbin}, {Wycoff}, {Bastian}, {Schwekendiek}, \&
  {Wicenec}}]{HogEtal2000aaTycho2}
{H{\o}g}, E. {et~al.} 2000, \aap, 355, L27

\bibitem[{{Horne}(1986)}]{Horne1986paspoptextr}
{Horne}, K. 1986, \pasp, 98, 609

\bibitem[{{Hubbard} {et~al.}(1997){Hubbard}, {Porco}, {Hunten}, {Rieke},
  {Rieke}, {McCarthy}, {Haemmerle}, {Haller}, {McLeod}, {Lebofsky},
  {Marcialis}, {Holberg}, {Landau}, {Carrasco}, {Elias}, {Buie}, {Dunham},
  {Persson}, {Boroson}, {West}, {French}, {Harrington}, {Elliot}, {Forrest},
  {Pipher}, {Stover}, {Brahic}, \& {Grenier}}]{HubbardEtal1997icsatmes}
{Hubbard}, W.~B. {et~al.} 1997, Icarus, 130, 404

\bibitem[{{Lang}(1992)}]{Lang1992bookADPS}
{Lang}, K.~R. 1992, {Astrophysical Data I.~Planets and Stars} (Springer-Verlag)

\bibitem[{{Shure} {et~al.}(1994){Shure}, {Toomey}, {Rayner}, {Onaka},
  {Denault}, {Stahlberger}, {Watanabe}, {Criez}, {Robertson}, \&
  {Cook}}]{ShureEtal1994exansfcam}
{Shure}, M. {et~al.} 1994, Exp.\ Astron., 3, 239

\bibitem[{{Skrutskie} {et~al.}(2006){Skrutskie}, {Cutri}, {Stiening},
  {Weinberg}, {Schneider}, {Carpenter}, {Beichman}, {Capps}, {Chester},
  {Elias}, {Huchra}, {Liebert}, {Lonsdale}, {Monet}, {Price}, {Seitzer},
  {Jarrett}, {Kirkpatrick}, {Gizis}, {Howard}, {Evans}, {Fowler}, {Fullmer},
  {Hurt}, {Light}, {Kopan}, {Marsh}, {McCallon}, {Tam}, {Van Dyk}, \&
  {Wheelock}}]{SkrutskieEtal2006AJ2MASS}
{Skrutskie}, M.~F. {et~al.} 2006, \aj, 131, 1163

\end{thebibliography}

\end{document}